\documentclass[aps,preprint,nofootinbib,12pt]{revtex4}
\usepackage{graphicx}
\usepackage{epsfig}
\usepackage[dvips]{color}
\begin{document}

\title{The Global Star Formation Law of Galaxies Revisited in the Radio Continuum}

\author{Lijie Liu$^{1,2}$, Yu Gao$^{1}$\footnote{yugao@pmo.ac.cn}}

\affiliation{
$^{1}$ {\footnotesize Purple Mountain Observatory, Chinese Academy of Sciences, China} \\
$^{2}$ {\footnotesize Graduate School of Chinese Academy of Sciences, China} \\
}

\begin{abstract} 
We study the global star formation law - the relation between the gas and star formation rate (SFR) in a sample of 130 local galaxies with infrared (IR) luminosities spanning over three orders of magnitude (${\rm 10^9-10^{12} L_\odot}$), 
which includes 91 normal spiral galaxies and 39 (ultra)luminous IR galaxies [(U)LIRGs]. We derive their total (atomic and molecular) gas and dense molecular gas masses using newly available HI, CO and HCN data from the literature. The SFR of galaxies is determined from total IR (${\rm 8-1000~\mu m}$) and 1.4 GHz radio continuum (RC) luminosities. The galaxy disk sizes are defined by the de-convolved elliptical Gaussian FWHM of the RC maps. 
We derive the galaxy disk-averaged SFRs and various gas
surface densities, and investigate their relationship. We find that the galaxy disk-averaged surface densities of dense molecular gas mass has the tightest correlation with that of SFR (scatter $\sim$ 0.26 dex), and is linear in $\log-\log$ space (power-law slope of $N=1.03 \pm
0.02$) across the full galaxy sample. The correlation between the total gas and
SFR surface densities for the full sample has a somewhat larger scatter ($\sim$ 0.48 dex), 
and is best fit by a power-law with slope $1.45 \pm 0.02$. However, the slope changes from $\sim 1$ when only normal spirals are considered, to $\sim1.5$ when more and more (U)LIRGs are included in the fitting. When different CO-to-${\rm H_2}$ conversion factors are used to infer molecular gas masses for normal galaxies and (U)LIRGs, the bi-modal relations claimed recently in CO observations of high-redshift galaxies appear to also exist in local populations of star-forming galaxies.  
\end{abstract}

\maketitle

\section{Introduction}

Understanding how stars form in galaxies is fundamentally one of the
central questions in galactic and extragalactic astronomy, 
since star formation (SF) plays an integral part of the formation and evolution of galaxies. 
Studies of the so-called SF law, which relates the SF rate
(SFR) to the density of the gas from which the stars form, have proven 
to be a fruitful way of approaching this problem (e.g., 
Kennicutt 1998, hereafter K98).

Schmidt (1959) first introduced the notion of a general SF law 
by suggesting that the SFR volume density ($\rho_{\rm {SFR}}$) varies with 
the interstellar gas volume density ($\rho_{\rm {gas}}$) to the power n, 
i.e.  $\rho _ {\rm {SFR}} \propto \rho _ {\rm {gas}} ^ {n}$, where $n \sim 2$. 
Under the assumption of a constant scale-height for the star-forming gas, 
this relation translates directly into an equivalent expression for 
the SFR and gas surface densities:  
$\Sigma _ {\rm {SFR}} \propto \Sigma _ {\rm {gas}} ^ {N}$. 
K98 studied the global SF law in log surface densities space and found that 
a power-law index of $N \sim 1.4 \pm 0.2$ provided a surprisingly tight parameterization of the data.
These studies of the Kennicutt-Schmidt SF law (KS law)
were among the first to highlight the importance of the gas density in regulating the SFR in star-forming galaxies, 
including luminous and ultraluminous infrared (IR) galaxies (hereafter (U)LIRGs, $L_{IR}\geq 10^{11}L_\odot$, Sanders \& Mirabel 1996).

A critical aspect of better determining the SF law is to accurately measure the physical area of a galaxy disk that is actively forming stars. 
In K98, where a total of more than 90 local galaxies were studied -- including 61 normal spiral disk galaxies, 
and 36 IR-selected starburst galaxies (normal star-forming galaxies with circumnuclear IR starbursts and (U)LIRGs) --
the sizes used to normalize the total SFR and gas masses for the subset of normal galaxies
were determined from optical images, while the sizes of IR starbursts were determined from CO or IR maps. 
Furthermore, different SFR estimators were employed by K98 in deriving the SFR of the various sub-samples. 
For normal galaxies, the SFR was derived from the H${\alpha}$ data, whereas 
for IR-starburst galaxies, 
the SFR was derived from their far-IR luminosities. 

In the work presented here, we attempt to establish the SF law based on 
the exact same observables for our entire galaxy sample. 
In particular, for every galaxy in our sample, we use high-resolution 
radio continuum (RC) maps to infer its star-forming disk size,  
and derive its SFR from both RC and IR luminosities. 
The RC emission at 1.4GHz is an excellent tracer of SFR, 
with the well-known far-IR-radio correlation being astonishingly tight, having only $\sim 0.3$ dex scatter 
over five orders of magnitude in luminosity (Condon et al. 1991; Yun et al. 2001; Murphy et al. 2006a). 
Recent spatially resolved measurements also show that 
this linear correlation is arguably valid locally within galaxy 
disks, down to $\sim1$ kpc or even sub-kpc scales (Murphy et al. 2006a, b, 2008).
We therefore expect our radio-inferred galaxy sizes to accurately match the 
star-forming regions in our galaxies. 
In starburst galaxies in particular, we expect the RC maps to give 
more truthful pictures of the extent of the star-forming region 
than the mixture of CO or IR maps (K98, both in many cases suffer 
from the limited sensitivity and resolution), since the latter may trace 
diffuse gas that is not actively forming stars (Bothwell et al. 2010; Rujopakarn et al. 2011). 
The RC emission, on the other hand, probes synchrotron radiation from 
supernovae remnants and thermal emission from HII regions, 
both of which have short-lived massive stars as progenitors, and is therefore likely to be a better tracer of the current SFR 
and its spatial extent of star-forming regions in galaxies.   

Millimeter and far-IR observations of molecular clouds in our Galaxy have shown that stars form in dense cores, 
and that massive stars form almost exclusively in clumps/clusters of massive dense cores (Evans 1999, 2008; Wu et al. 2010). 
The dense gas ($n > 10^4 {\rm cm}^{-3}$) residing in these cores is best traced by molecules
with a high-dipole moment requiring high critical densities, such as HCN and CS.
In fact, Gao \& Solomon (2004a, b, hereafter, GS04a, b) carried out 
one of the first systematic HCN surveys of a large local galaxy sample and
found a tight linear correlation between the IR and HCN luminosities spanning three orders of magnitude. 
More recent studies, targeting not just HCN but also other high-density gas 
tracers (e.g. {\rm HCO$^{+}$}, HNC, CN and CS) in local galaxies, 
also find molecular line luminosities linearly increasing with 
${\rm L_{IR}}$ (Baan et al. 2008; Krips et al. 2008; Juneau et al. 2009; 
Matsushita et al. 2010). 
Liu \& Gao (2010) further showed that RC and HCN correlate nearly
equally as well as that of the IR and HCN correlation (GS04a, b).
 
Therefore, we use the uniformly calibrated SFR and 
the consistently derived well-matched galaxy size measurements from the radio maps in all 
star-forming galaxies to present an analysis of the dense gas SF law 
based on HCN observations in addition to the
traditional KS law of total gas in this paper. Using both samples of 
K98 and GS04b, we scale both the total gas and dense 
molecular gas masses using the measured RC sizes to study the SF 
law in terms of dense molecular gas (${\rm \Sigma_{SFR}}$ versus 
${\rm \Sigma _{dense}}$) as well as the traditional KS law in terms 
of the total gas mass. Further details in comparing the surface SFR with 
surface densities of the total gas, molecular gas and dense molecular
gas to derive the various forms of the SF law, the comparison among
them, and various related issues will be discussed with an even larger
sample of galaxies in a
future paper (Liu, Gao \& Greve 2011, in preparation).

\section{Sample and Data}

Our sample includes all galaxies from both K98 and GS04b, totaling 
130 galaxies with IR luminosities spanning 3 orders of
magnitude (${\rm 10^9 - 10^{12} L_\odot}$), of which 39 are (U)LIRGs. 
The resulting sample is one of the largest samples of local star-forming galaxies with HI, CO and RC measurements available.
Furthermore, HCN measurements are available for about half of our sample sources. The distances of these galaxies come from
the  on-line CDS data base. We adopt a Hubble constant of $H_0 = 73~{\rm km~s^{-1}~Mpc^{-2}}$. 

Normal disk galaxies of K98 were mainly selected from the FCRAO CO 
survey (Young et al. 1989, 1995) and CO survey of Sage (1993) with published HI maps and H$\alpha$ photometry at that time.  
We derive the total atomic gas mass mainly from the HyperLeda homogenized HI catalogue (Paturel et al. 2003), 
supplemented by some other individual measurements. CO data are gathered from single-dish 
maps/observations of galaxies (e.g., Kuno et al. 2007; 
Chung et al. 2009). Most of our HCN data come 
from the systematic survey of GS04a, with a few from sparse 
sampling of the other papers (Baan et al. 2008; Krips et al. 2008; 
Graci\'a-Carpio et al. 2008a; Juneau et al. 2009).

We derived the SFR from both RC and IR luminosities with the calibrations from Bell (2003).
The IR (${\rm 8 -1000~\mu m}$) luminosities are calculated using the fluxes in all four IRAS bands 
which are taken mainly from Sanders et al. (2003).
Integrated 1.4GHz RC fluxes were taken mainly from the systematic 
NRAO VLA sky survey (NVSS, Condon et al. 1998; Condon et al. 2002). 
The 20cm RC images are used to measure the
physical sizes of all galaxies.  They are defined as the de-convolved 
FWHM of the fitted Gaussian ellipticals. For some galaxies without 
suitable radio maps from literature, we reduced the NRAO VLA archival data 
and made the RC maps by ourselves. More details can be found
in Liu (2011) and will be 
given in a future paper (Liu et al. 2011, in prep.). 

\section{Results and Analysis}

As mentioned above, the SFR was derived in two independent ways using the RC and IR data. 
Excellent agreement was found between the two SFR estimators, and using 
one over the other, therefore, did not alter our results in any 
significant way (Liu et al., in prep.).
For the remainder of this paper we only adopt the IR-based SFR.

\subsection{${\rm \Sigma_{SFR}}$ versus $\Sigma$gas}
Figure 1a shows the relationship between $\Sigma$gas and ${\rm \Sigma_{SFR}}$,
where the Galactic CO-to-${\rm H_2}$ conversion factor of  $4.6~{\rm M_\odot~ (K~Km/s~pc^2)^{-1}}$
(e.g. Solomon et al. 1987) has been used to infer ${\rm H_2}$ mass for 
the entire sample. 
A strong correlation between total gas and SFR surface densities is found with 
a least squares best fitting slope of ${\rm N=1.45 \pm 0.02}$, 
which is in excellent agreement with the slope (${\rm N =1.4  \pm 0.15}$) found by K98.
However, we find the slope is nearly one for normal disk spirals and
changes from 1 to 1.5 depending upon how many (U)LIRGs are included 
in the total sample of galaxies (similar to Fig. 7 in GS04b). 

A likely caveat of the above findings, however, is introduced by the mismatch between the spatial scales 
of the HI, CO, IR (SFR) and RC emission. 
HI data are from single-dish measurements, encompassing mostly the 
entire HI gas reservoir, 
which is often found to extend well beyond the optical (and star-forming) sizes of galaxies (e.g. Garc{\'{\i}}a-Ruiz et al. 2002).
In particular, it is true that significantly extended HI gas is 
distributed well beyond the molecular disk for normal spiral galaxies 
(often
comparable in mass to H$_2$ or even dominant in total gas), whereas 
the bulk of the total gas is in molecular form for most (U)LIRGs 
with most of HI in tidal debris. 
In contrast, the star-forming areas derived from high-resolution RC measurements, which are used to infer the gas and SFR surface densities,  
will necessarily be significantly smaller than the area probed by the HI observations for galaxies, particularly for (U)LIRGs.

\begin{figure}
  \centering
  \includegraphics[width=0.85\textwidth]{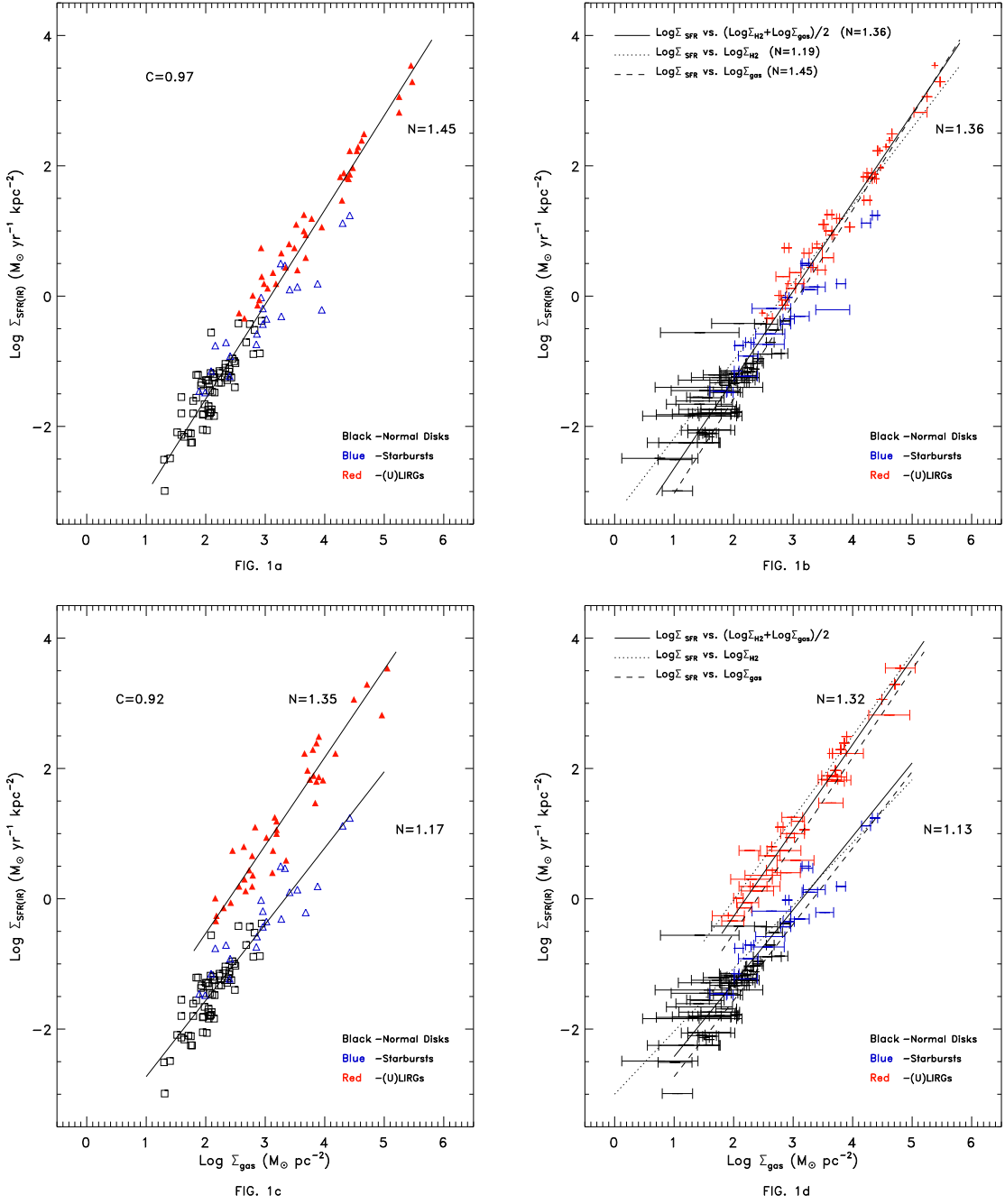}
  \caption{ {\footnotesize The correlations between ${\rm \Sigma_{SFR}}$ and ${\rm \Sigma_{gas}}$, with normal disks,
   star-forming galaxies with circumnuclear starbursts (starbursts) and (U)LIRGs  shown in black, blue and red symbols, respectively.
    (a) With the same  $\alpha_{\rm CO}$ ($\sim 4.6$) for all galaxies, 
a slope of ${\rm N=1.45 \pm 0.02}$ is derived. 
  (b) The ranges of total gas surface densities are shown as the
gas surface densities plotted in (a) are strictly-speaking upper limits (see text). 
  The dashed, dotted and solid lines are least squares fits to the lower and upper limits and average values [${\rm (Log\Sigma_{H2}+Log\Sigma_{gas}})/2$], with slopes of $1.19\pm0.03$, $1.45\pm0.02$ and $1.36\pm0.02$, respectively.   
(c) With different conversion factors for (U)LIRGs 
[$\alpha_{\rm CO} \sim 0.8$ as compared to $\alpha_{\rm CO} \sim 4.6$ used 
in (a)], the fitting lines for normal disks and (U)LIRGs have slopes of $1.17\pm0.05$ and $1.35\pm0.06$, respectively. 
(d) Different $\alpha_{\rm CO}$ and the ranges of total gas densities.  
The dotted, dashed and solid lines are least squares fits to the lower and upper limits and 
average values of gas surface densities, respectively yielding slopes of 
$0.97\pm0.05$, $1.17\pm0.05$ and $1.13\pm0.05$ for normal galaxies and
$1.26\pm0.04$, $1.35\pm0.06$ and $1.32\pm0.04$ for (U)LIRGs. Fitting the entire sample can lead to a slope as high as $\sim 1.8$.} }
\end{figure}

As a result, by scaling the global HI mass with our measured radio 
disk sizes, 
we will very likely overestimate the disk-averaged atomic (and thus the total) gas surface densities, 
and these should therefore be better regarded as the upper limits.
At the same time, the ${\rm H_2}$ surface densities can be regarded as strictly the lower limits on the total gas surface densities 
(provided, of course, that the CO-to-${\rm H_2}$ conversion factor applied is appropriate).
In Figure 1b, we have, for each galaxy, indicated the possible ranges of the total gas surface densities,
with ${\rm \Sigma_{H2}}$ as the lower limits and ${\rm \Sigma_{gas}}$ calculated from the sum of the global HI measurements and ${\rm H_2}$ as the upper limits.
Separate least squares fits to these lower and upper limits yield slopes 
of  $1.19\pm0.03$ and $1.45\pm0.02$, respectively. 
Using the average value of these limits 
[${\rm (Log\Sigma_{H2}+Log\Sigma_{gas}})/2$], 
we obtain a slope of $1.36\pm0.02$. 

One can see that normal galaxies in our sample exhibit significantly larger ranges in ${\rm \Sigma_{gas}}$ 
than the (U)LIRGs, in line with our previous statement that normal galaxies tend to have much larger HI fractions.
In fact, in many of the low luminosity galaxies, the atomic gas content can dominate the total gas mass,
while for (U)LIRGs, the atomic gas is much smaller, 
and often negligible, compared with the molecular gas content. 
The higher fraction of atomic gas in low luminosity galaxies is probably 
part of the reason why the KS law of total gas has a higher slope ($\sim1.45$) than 
the KS law in terms of molecular gas ($\sim 1.19$). 

Another thorny issue is the CO-to-${\rm H_2}$  conversion factor 
used in deriving molecular gas masses.
K98 used the same Galactic value ($ \alpha_{CO}=4.6~{\rm M_{\odot}~(K~km~s^{-1}~pc^2)^{-1}}$) 
for all objects in his sample, as indeed what we have adopted above. 
However,  CO studies of (U)LIRGs, both near and far, strongly argue in favor 
of a substantially smaller conversion factor for (U)LIRGs, by 
about a factor of 6 (Solomon et al. 1997; Downes \& Solomon 1998; 
Solomon \& Vanden Bout 2005; Bouch${\rm \acute{e}}$ et al. 2007).
In galaxy mergers, the physical properties of the molecular gas are strongly affected by the extreme environments. 
The rise in velocity dispersion and kinetic temperature in GMCs during the merger increases the CO intensity, 
and lowers the observed $\alpha co$ from the Galactic value by a typical factor of $\sim 2-10$ (Narayanan et al. 2011). 
In the following, we discuss the effects of using these different
CO-to-${\rm H_2}$ conversion factors on the SF laws.

In Figure 1c, we have used the more appropriate conversion factor of 
0.8 $M_\odot~ {\rm (K~Km/s~pc^2)^{-1}}$ for (U)LIRGs 
(but maintained the Galactic value for normal galaxies).
As a result, normal galaxies and (U)LIRGs split up into 
two distinct ${\rm \Sigma_{SFR}}$ versus ${\rm \Sigma_{gas}}$ relations
as compared to Fig. 1a. 
Fitting to normal galaxies and (U)LIRGs separately yields slopes of $1.17\pm0.05$ and $1.35\pm0.06$, respectively.
As before, considering that the gas surface densities here are merely 
upper limits, 
we show in Figure 1d the ranges of total gas densities with 
different $\alpha_{CO}$. 
The dotted, dashed and solid lines are least squares fits to the lower and upper limits and 
average values [${\rm (Log\Sigma_{H2}+Log\Sigma_{gas}})/2$] of gas surface densities, respectively, for each sub-sample.  
For the normal galaxies, fits to the lower and upper limits and average values yield 
$0.97\pm0.05$, $1.17\pm0.05$ and $1.13\pm0.05$, respectively.
While for (U)LIRGs, the same fits yield slopes of  $1.26\pm0.04$, $1.35\pm0.06$ and $1.32\pm0.04$, respectively.  Fitting the entire 
sample of both normal galaxies 
and (U)LIRGs can lead to a slope as high as $\sim 1.8$.

The two different ${\rm \Sigma_{SFR}}$ -- ${\rm \Sigma_{gas}}$ relations found in this manner, one for normal galaxies
and one for (U)LIRGs, appear to be same to the bi-modal relations recently found from high-z CO observations where two different CO-to-${\rm H_2}$ conversion factors, same as here, were used 
(Daddi et al. 2010; Genzel et al. 2010; Ivison et al. 2011).
In this picture, both local and high-z (U)LIRGs have $\sim 10$ times 
higher ${\rm \Sigma_{SFR}}$ at fixed gas density
compared to normal disk galaxies (both local and at high-z).
The slope for (U)LIRGs is also steeper than for normal galaxies, 
which suggests that the SF efficiency is higher in (U)LIRGs 
than in normal spirals.

\subsection{${\rm \Sigma_{SFR}}$ versus ${\rm \Sigma _{dense}}$}

We normalize the SFR and dense molecular gas mass using measured radio 
sizes to study the SF law in terms of dense molecular gas. 

\begin{figure}
\small
  \centering
  \includegraphics[width=0.85\textwidth]{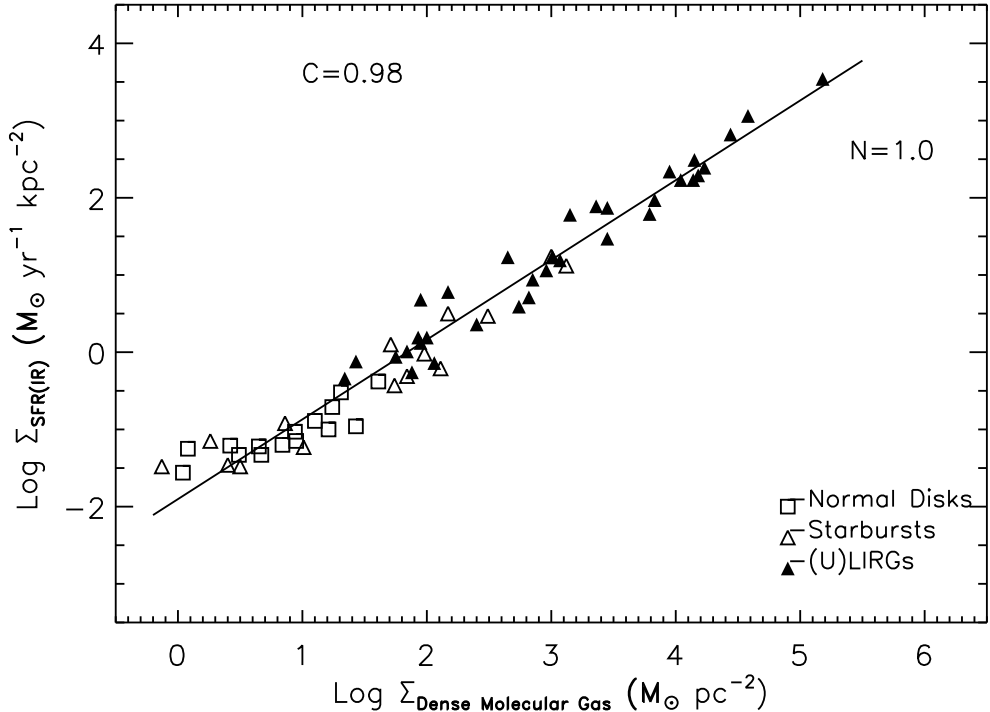}
  \caption{ {\footnotesize ${\rm \Sigma_{SFR}}$ vs. ${\rm \Sigma _{dense}}$ for all galaxies in the sample (open squares for normal disks, open triangles for normal starburst galaxies and solid triangles for (U)LIRGs). The relation is linear, with power law index $N=1.03 \pm 0.02$ and correlation coefficient 0.98.} }
\end{figure}

Figure 2 presents the derived ${\rm \Sigma _{dense}}$ and ${\rm \Sigma_{SFR}}$ for the 64 galaxies with HCN measurements,  
and a very tight correlation spanning more than 5 orders of magnitude in ${\rm \Sigma _{dense}}$ 
(the correlation coefficient C=0.98) is found. 
The dispersion in the relation is much smaller compared to those in Fig. 1, having about 50\% less
scatter in surface SFR density at a fixed surface dense molecular gas density.

A least squares fit to the data results in a power index $N=1.03 \pm 0.02$: 
\begin{equation}
{\rm \log \Sigma_{SFR} = (1.03 \pm 0.02) \log{\Sigma_{dense}} + (-1.90 \pm 0.07)}
\end{equation}

Therefore,  the SF law in terms of dense molecular gas has a power-law index of exactly 1.0, in agreement with the tight linear relation 
between ${\rm L_{IR}}$ and ${\rm L_{HCN}}$ (GS04a,b). 
Scaling the dense molecular gas mass and SFR by our measured RC disk 
sizes does not 
change the slope of the ${\rm L_{IR}}$-${\rm L_{HCN}}$ correlation 
since this relation is essentially exactly linear. 
Graci\'a-Carpio et al. (2008b) suggested a two-function power law in terms of dense molecular gas and SFR densities, 
which claimed a slightly higher slope in (U)LIRGs than in normal galaxies. 
We also fit the normal galaxies and (U)LIRGs separately and find that there is no significant difference in the slope. 
It turns out that both the power indexes are close to unity: 0.96 $\pm$ 0.05 for normal galaxies and 1.01 $\pm$ 0.03 for (U)LIRGs. 
Therefore, the tight linear correlation between ${\rm \Sigma _{dense}}$ 
and ${\rm \Sigma_{SFR}}$ 
is established within the whole sample of galaxies, independent of IR luminosity. 
This linear SF law in terms of dense molecular gas (Fig. 2) is drastically different from the usual KS law in terms of total 
gas since there are no unique slopes in the fits and the fitted 
slopes change according to the sample (Fig. 1).

\section{Discussion}   

Figs. 1 and 2 show very strong correlations between 
the gas and SFR surface densities, which suggests that gas density 
(most importantly, the dense gas) plays an important role in 
regulating the SF in galaxies. 
Although a power-law slope of $1.45 \pm 0.02$ is found for the entire sample in Fig. 1 --
the usual KS law in terms of total 
gas, the slopes change from nearly linear 
for normal spirals to 1.5 when starbursts and (U)LIRGs are included.
The slope of $N \sim 1.5$ is expected for self-gravitating gas disks,  
where the SFR scales as the ratio of the gas density ($\rho$) to the 
free-fall time scale ($\tau_{ff} \propto \rho^{-0.5}$) (Elmegreen 2002; Kennicutt et al. 2007).  

K98 divided the sample into normal spiral disk galaxies and IR-selected starburst galaxies, 
with the latter including both normal star-forming disk galaxies with 
circumnuclear starbursts and (U)LIRGs. However, at least five same spiral disk galaxies 
(NGC 2903, 4736, 5194, 5236, 6946) are in both sub-samples of 
normal spiral disk galaxies and IR-selected starburst galaxies 
in K98, and 
certainly quite a few more starburst galaxies could easily be
attributed to normal spirals as well (such as NGC 891).
The problem here is that this definition would affect the 
adopted ways to derive the SFR and size of a galaxy, and thus, would affect 
the final fitting results of the SF law, 
since K98 used different ways to derive the SFR and galaxy sizes for 
these two sub-samples.

The differences in the galaxy size measurements directly affect the 
derived surface SFR and gas surface densities. This is particularly true 
for the dust-obscured starbursts and (U)LIRGs, where optical sizes may not
reflect the regions of active SF. Instead, interferometric CO maps have been used
for some starbursts and (U)LIRGs, yet these are often limited in spatial extent 
and sensitivity (due to poor uv coverage and limited integration time). 
As a result, they may not accurately trace the sizes of active SF either.

The strategy adopted here is to use the sensitive high-resolution 1.4\,GHz RC maps
to infer the sizes of the star-forming areas in galaxies. The philosophy behind is that RC emission is a well-known tracer of SF activity 
(Condon et al. 1991; Yun et al. 2001; Murphy et al. 2006a, b).
Comparing our measured radio sizes 
with CO sizes from the literature for most of the starburst sample, 
we find good agreement between the two size estimators. 
This confirms the findings of other studies 
of local (U)LIRGs, where the radio emission has also been found to mimic that of CO emission
(Downes \& Solomon 1998; Tacconi et al. 1999; Iono et al. 2007; Bouche et al. 2007). 
 It seems, therefore, that for starbursts and (U)LIRGs at least, we can confidently
use the radio size  as a tracer of the global distribution of SF and molecular gas.

Optical sizes of galaxies are also compared with radio sizes in this work. 
For the normal galaxies, we find that the ratios of 
optical sizes to radio sizes show relatively small scatter, and they are 
in a range of 2--5, with an average ratio of $\sim 3.3$. 
The ratios for those normal active galaxies which are known to contain 
an AGN are, not surprisingly, somewhat higher, 
but the difference is still marginal since the average ratio is 
$\sim 4.1$. 
Thus, normalizing the SFRs and molecular gas masses of the normal galaxies
 with optical sizes (cd. K98) would tend to underestimate the 
surface densities, possibly by an order of magnitude, compared to the 
normalization inferred from the radio sizes. This size difference might
help us better understand the major contributing factors to the poor 
correlations found in normal spirals (Figs. 2 and 4 in K98).

In addition, we find that the KS law for (U)LIRGs does not only exhibit
a smaller scatter ($\rm \sim 0.31~dex$ vs. 0.48~dex for entire sample), 
but also has a higher slope than the
case for the normal galaxies. In fact, the slope changes from $\sim 1.0$ to $\sim 1.8$ with different conversion factors 
when more and more (U)LIRGs are included. 
This may imply a density-dependent SF efficiency of gas, 
i.e., the higher the density, the higher SF efficiency. 
A similar trend was also pointed out by GS04b, who studied
the global IR-CO luminosity correlation of local (U)LIRGs.
Studies of high-$z$ galaxies indicate a similar behavior
(Bouch${\rm \acute{e}}$ et al. 2007; Daddi et al. 2010; Bothwell et al. 2010). 

With a smaller conversion factor for the (U)LIRGs, 
an even higher slope in KS law is expected. 
The SF efficiency of these (U)LIRGs will be even higher given the smaller
molecular gas masses. 
The bi-modal relation in KS law, more obviously shown with different conversion factors in the local
sample of galaxies, is similar to those recently claimed
in high redshift CO observations (Daddi et al. 2010; Genzel et al. 2010). 
One can easily see (Fig. 1c, d) that the two different SF 
regimes suggested by the high redshift works also exist in the local universe: 
a long-lasting mode for normal disks and a more rapid mode for 
starbursts/(U)LIRGs.  

Considering the atomic gas mass is derived from global HI measurements, 
we then have, for each galaxy, indicated the possible range of its 
total gas surface density, 
with ${\rm \Sigma_{H2}}$ as the lower limit and ${\rm \Sigma_{gas}}$ calculated
 from global HI measurements as the upper limit.
We find that normal galaxies in our sample exhibit significantly larger ranges in ${\rm \Sigma_{gas}}$ 
than the (U)LIRGs. This is consistent with the fact that normal galaxies tend to have larger HI fractions than (U)LIRGs. Nevertheless, the changes
in various fit slopes from these 
limits are small, and the slope changes more dramatically when the sample of galaxies 
changes by the inclusion of more and more (U)LIRGs.

In contrast to the total gas, the surface SF efficiency of dense molecular gas
(${\rm \Sigma_{SFR}}$/${\rm \Sigma _{dense}}$) is constant in all galaxies
regardless of the galaxy luminosity.  ${\rm \Sigma _{dense}}$ shows the best
relation with ${\rm \Sigma_{SFR}}$, which is linear with index $N=1.03
\pm 0.02$.  Fitting the normal galaxies and (U)LIRGs separately yields the same
linear slope, and the scatter is also much smaller ($\sim 0.26$~dex).  Therefore, the SF law relating to the dense gas is not only
linear and much tighter, as compared to the usual KS law relating to the
total gas, it is also seemingly universal across
galaxy types. This is consistent with the linear
$\rm L_{IR} - L_{HCN}$ correlation amongst local normal galaxies and 
(U)LIRGs (GS04b) and almost linear correlation even in high-z galaxies (Gao et al. 2007).
An HCN survey of Galactic dense cores by Wu et al.\ (2005) extended this linear
luminosity correlation down to much smaller scales.  Both GS04b and Wu et al.\
(2005) explain this tight, linear luminosity relation in terms of the dense
cores being the basic units of SF in galaxies.  In this scenario,
${\rm \Sigma _{dense}}$ merely reflects how many such dense cores exist on
average per unit area in a galaxy, in which case a linear correlation between
${\rm \Sigma_{SFR}}$ and ${\rm \Sigma _{dense}}$ on global galaxy scale
naturally arises. Hence this tight linear relation has little or no dependence
upon the galaxy luminosity.

\section{SUMMARY}
We have re-examined the global star formation (SF) law -- the relation between 
the galaxy-averaged surface densities of the gas (HI and ${\rm H_2}$) and SF rate (SFR)
in a sample of 130 local galaxies with IR luminosities spanning three orders of magnitude (${\rm 10^9 -10^{12} L_\odot}$),
which includes 91 normal spiral galaxies and 39 (U)LIRGs. 
Unlike previous studies, which have used optical, IR or CO observations to
infer the galaxy sizes of star-forming areas, we have taken a novel
approach and used high-resolution radio continuum observations to uniformly
measure the radio sizes for all galaxies in the sample.  Accurate size
determination of the area of active SF within the galaxies is a key step, as it
directly affects the inferred surface densities for gas and SFR.

We recover the KS law between total gas and SFR surface densities, finding a slope of $1.45 \pm 0.02$.
The (U)LIRG population, however, appear to show a tighter and steeper relation than normal galaxies, 
which implies a density-dependent SF efficiency of total gas. 
 In fact, we find the slope of the SF law changes from 1.0 to 1.5 
 depending upon how many (U)LIRGs are in the sample.
The SF law shows the same bi-modal relations
claimed recently in high redshift CO observations, 
suggesting the two different SF regimes also exist in the local universe. 
This is particularly true when a different CO-to-${\rm H_2}$ conversion factor is used for (U)LIRGs, leading to even higher SF efficiency 
in (U)LIRGs. 
Dense molecular gas surface density shows the best correlation with SFR surface density, 
which is exactly linear with index $N=1.03 \pm 0.02$.  
Fitting the normal galaxies and (U)LIRGs separately yields the same linear slope,   
which suggests the SF efficiency of dense molecular gas is the same in all types of galaxies, regardless of galaxy luminosity. 
This tight relation can be explained if the basic units of  SF in galaxies are in dense cores.

\section*{Acknowledgments}

We are grateful to Dr. Thomas R.\ Greve for 
his constructive suggestions and advices. This work was supported by
the National Natural Science Foundation of China (Grant Nos.   
10833006 and 10621303), 
and the National Basic Research Program of China (Grant No. 2007CB815406).
This research had made use of the NASA/IPAC Extragalactic Database and NRAO Science Data Archive.


\begin{thebibliography}{}
\bibitem{Bell2003} Bell, E.~F.\ 2003, ApJ, 586, 794 
\bibitem{Baan2008} Baan, W.~A., Henkel, C., Loenen, A.~F., Baudry, A., \& Wiklind, T.\ 2008, A\&A, 477, 747 
\bibitem{Bouche} Bouch{\'e}, N., et al.\ 2007, ApJ, 671, 303 
\bibitem{Bothwell2010} Bothwell, M.~S., et al.\ 2010, MNRAS, 405, 219
\bibitem{Condon1991} Condon, J.~J., Anderson, M.~L., \& Helou,~G.\ 1991, ApJ, 376, 95
\bibitem{Condon et al.(1998)} Condon, J.~J., Cotton, W.~D., Greisen, E.~W., Yin, Q.~F., Perley, R.~A., Taylor, G.~B., \& Broderick, J.~J.\ 1998, AJ, 115, 1693
\bibitem{Condon2002} Condon, J.~J., Cotton, W.~D., \& Broderick, J.~J.\ 2002, AJ, 124, 675  
\bibitem{Chung2009} Chung, E.~J., Rhee, M.-H., Kim, H., Yun, M.~S., Heyer, M., \& Young, J.~S.\ 2009, ApJS, 184, 199 
\bibitem{DS1998} Downes, D., \& Solomon, P.~M.\ 1998, ApJ, 507, 615 
\bibitem{Daddi2010} Daddi, E., et al.\ 2010, ApJ, 714, L118
\bibitem{Elmegreen(2002)} Elmegreen, B.~G.\ 2002, ApJ, 577, 206
\bibitem{Evans(1999)} Evans, N.~J., II 1999, ARA\&A, 37, 311
\bibitem{Evans2008)} Evans, N.~J., II 2008, Pathways Through an Eclectic Universe, 390, 52 
\bibitem{GarciaRuiz2002} Garc{\'{\i}}a-Ruiz, I., Sancisi, R., \& Kuijken, K.\ 2002, A\&A, 394, 769 
\bibitem{GC2008a} Graci{\'a}-Carpio, J., Garc{\'{\i}}a-Burillo, S., \& Planesas, P.\ 2008a, Ap\&SS, 313, 331 
\bibitem{GC2008b} Graci{\'a}-Carpio, J., Garc{\'{\i}}a-Burillo, S., Planesas, P., Fuente, A., \& Usero, A.\ 2008b, A\&A, 479, 703 
\bibitem{Genzel et al.(2010)}Genzel, R., et al.\ 2010, MNRAS, 407, 2091
\bibitem{GS04a} Gao, Y., \& Solomon, P.~M.\ 2004a, ApJS, 152, 63 
\bibitem{GS04b} Gao, Y., \& Solomon, P.~M.\ 2004b, ApJ, 606, 271
\bibitem{Gao et al.(2007)} Gao, Y., Carilli, C.~L., Solomon, P.~M., \& Vanden Bout, P.~A.\ 2007, ApJ, 660, L93 
\bibitem{Iono et al. 2007}Iono D., Wilson C.~D., Takakuwa S., Yun M.~S., Petitpas G.~R., Peck A.~B., Ho P. T.~P., Matsushita S., et al.,\ 2007, ApJ, 659, 283
\bibitem{Ivison et al.(2011)} Ivison, R.~J., Papadopoulos, P.~P., Smail, I., Greve, T.~R., Thomson, A.~P., Xilouris, E.~M., \& Chapman, S.~C.\ 2011, MNRAS, 412, 1913
\bibitem{Juneau2009} Juneau, S., Narayanan, D.~T., Moustakas, J., Shirley, Y.~L., Bussmann, R.~S., Kennicutt, R.~C., \& Vanden Bout, P.~A.\ 2009, ApJ, 707, 1217 
\bibitem{K98} Kennicutt, R.~C.\ 1998, ApJ, 498, 541
\bibitem{Kennicutt et al.(2007)} Kennicutt, R.~C., Jr., et al.\ 2007, ApJ, 671, 333
\bibitem{Krips2008} Krips, M., Neri, R., Garc{\'{\i}}a-Burillo, S., Mart{\'{\i}}n, S., Combes, F., Graci{\'a}-Carpio, J., \& Eckart, A.\ 2008, ApJ, 677, 262 
\bibitem{Kuno2007} Kuno, N., et al.\ 2007, PASJ, 59, 117 
\bibitem{LiuGao(2010)} Liu, F., \& Gao, Y.\ 2010, ApJ, 713, 524
\bibitem{Matsushita2010} Matsushita, S., Kawabe, R., Kohno, K., Tosaki, T., \& Vila-Vilar{\'o}, B.\ 2010, PASJ, 62, 409 
\bibitem{Murphy2006a} Murphy, E.~J., et al.\ 2006a, ApJ, 638, 157 
\bibitem{Murphy et al.(2006)b} Murphy, E.~J., et al.\ 2006b, ApJ, 651, L111 
\bibitem{Murphy et al.(2008)} Murphy, E.~J., Helou, G., Kenney, J.~D.~P., Armus, L., \& Braun, R.\ 2008, ApJ, 678, 828 
\bibitem{Narayanan et al.(2011)} Narayanan, D., Krumholz, M., Ostriker, E.~C., \& Hernquist, L.\ 2011, arXiv:1104.4118
\bibitem{Paturel2003} Paturel, G., Theureau, G., Bottinelli, L., Gouguenheim, L., Coudreau-Durand, N., Hallet, N., \& Petit, C.\ 2003, A\&A, 412, 57
\bibitem{Rujopakarn et al.(2011)} Rujopakarn, W., Rieke, G.~H., Eisenstein, D.~J., \& Juneau, S.\ 2011, ApJ, 726, 93
\bibitem{SandersMirabel(1996)} Sanders, D.~B., \& Mirabel, I.~F.\ 1996, ARA\&A, 34, 749 
\bibitem{Sage(1993)} Sage, L.~J.\ 1993, A\&AS, 100, 537
\bibitem{Schmidt1959} Schmidt,~M.\ 1959, ApJ, 129, 243
\bibitem{Solomon et al.(1987)} Solomon, P.~M., Rivolo, A.~R., Barrett, J., \& Yahil, A.\ 1987, ApJ, 319, 730 
\bibitem{Solomon et al.(1997)} Solomon, P.~M., Downes, D., Radford, S.~J.~E., \& Barrett, J.~W.\ 1997, ApJ, 478, 144 
\bibitem{SolomonVandenBout2005} Solomon, P.~M., \& Vanden Bout, P.~A.\ 2005, ARA\&A, 43, 677 
\bibitem{Sanders2003} Sanders, D.~B., Mazzarella, J.~M., Kim, D.-C., Surace, J.~A., \& Soifer, B.~T.\ 2003, AJ, 126, 1607 
\bibitem{Tacconi et al. 1999}Tacconi L.~J., Genzel R., Tecza M., Gallimore J.~F., Downes D., Scoville N.~Z.,\ 1999, ApJ, 524, 732
\bibitem{Wu2005} Wu, J., Evans, N.~J., II, Gao, Y., Solomon, P.~M., Shirley, Y.~L., \& Vanden Bout, P.~A.\ 2005, ApJ, 635, L173 
\bibitem{Wu et al.(2010)} Wu, J., Evans, N.~J., Shirley, Y.~L., \& Knez, C.\ 2010, ApJS, 188, 313 
\bibitem{Young et al.(1989)} Young, J.~S., Xie, S., Kenney, J.~D.~P., \& Rice, W.~L.\ 1989, ApJS, 70, 699
\bibitem{Young et al.(1995)} Young, J.~S., et al.\ 1995, ApJS, 98, 219
\bibitem{Yun2001} Yun, M.~S., Reddy, N.~A., \& Condon, J.~J.\ 2001, ApJ, 554, 803 
\end{thebibliography}
\end{document}